\documentclass{cernrep}
\begin{document}
\title{Incorporating Nuisance Parameters in Likelihoods for Multisource Spectra}
\author{J. S. Conway}
\institute{University of California, Davis, USA}
\maketitle

\begin{abstract}
We describe here the general mathematical approach to constructing
likelihoods for fitting observed spectra in one or more dimensions
with multiple sources, including the effects of systematic
uncertainties represented as nuisance parameters, when the likelihood
is to be maximized with respect to these parameters.  We consider
three types of nuisance parameters: simple multiplicative factors,
source spectra ``morphing'' parameters, and parameters representing
statistical uncertainties in the predicted source spectra.
\end{abstract}

\section{Overview}

In particle physics one often encounters the general problem of
estimating physical parameters such as particle masses or cross
sections from the spectra of observables calculated in each event.  In
the case of a known, well-established signal process, the dominant
technique by far is to use a binned likelihood assuming a Poisson
distribution in each bin~\cite{Cowan}, and find the parameters which
maximize the likelihood.  

In the case of a search for a new particle or effect resulting in
either a discovery or null result, binned likelihoods have also been
employed successfully to quote statistical significance or exclusion
bounds, respectively.  From a certain point of view there is a
desirable consistency in utilizing the same basic statistical method
for searches, discoveries, and measurements.

A key requirement here, however, is that the likelihood somehow
incorporate the effects of all systematic uncertainties present in the
analysis.  In frequentist methods this is almost always achieved by
generating distributions of many pseudoexperiments, where from one
pseudoexperiment to the next the values of all parameters are varied
within their assumed distributions.  In a formal Bayesian treatment,
the nuisance parameters are removed by marginalization: integrating
them out, assuming some prior pdf.  Both of these approaches are
computationally very expensive.

In measuring parameters using binned Poisson likelihoods, as mentioned
above, one simply maximizes the likelihood (in practice one minimizes
the negative log of the likelihood) with respect to all $m$ free
parameters, and then constructs the standard error ellipsoid in
$m$-dimensional space.  The fit values of the nuisance parameters are
typically of no interest, leaving one to interpret the intervals for
just the parameters of interest in a straightforward way~\cite{pdg}.

We define in this paper three main types of nuisance parameters 
representing systematic uncertainties on the source distributions, and
describe how to incorporate them into a binned Poisson likelihood.

We further argue in this paper that this maximum likelihood method,
also called the profile likelihood, can be applied to searches and
discoveries as well, either by a pseudo-Bayesian interpretation of the
profile likelihood as representing a posterior density in the
parameter(s) of interest, or by likelihood ratio methods.  The profile
likelihood requires significantly less computer time, often a much as
two orders of magnitude less, than frequentist or frequentist-inspired
methods such as CL$_{\rm s}$~\cite{CLs}.  That in turn allows much
more detailed study of the properties of the fit results.

\section{Core of the Poisson Likelihood}

Suppose we observe in a set of $N$ events an observable or in general
a set of observables $\bar{x}$.  If we define a set of $n_{bin}$ bins
(which can be of literally any shape we choose) in the space of the
observables, then the number of events $n_i$ in each bin $i$, is 
assumed to be Poisson-distributed according to 
\begin{eqnarray}
     {\cal P}(n_i|\mu_i) = \frac{\mu_i^{n_i} e^{-\mu_i}}{n_i!}
\end{eqnarray}
where $\mu_i$ is the number of expected events in the bin.  Typically
we can write 
\begin{eqnarray}
    \mu_i = \sum_{j=1}^{n_{source}} L \sigma_j \epsilon_{ji}
\end{eqnarray}
for integrated luminosity L, cross section $\sigma_j$ for source $j$,
and efficiency $\epsilon_{ji}$ for source $j$ in bin $i$, often
obtained from MC simulation of the process.  The sources here include
the signal process of interest and all background processes.  Again,
since we are dealing with a possibly multidimensional space of
observables, the index $i$ can actually label the bins in multiple
dimensions.

The Poisson likelihood for the full observed spectrum is simply the 
product of the Poisson probabilities:
\begin{eqnarray}
    {\cal L} = \prod_{i=1}^N {\cal P}(n_i|\mu_i) \ \ \ .
\end{eqnarray}
In the absence of any systematic uncertainties one can simply minimize
$-\ln{\cal L}$ with respect to all unknown parameters in the problem and
interpret the resulting standard error ellipsoid in the normal way to 
obtain estimates of the unknown parameters and associated confidence
intervals.

\section{Multiplicative Uncertainties}

Multiplicative uncertainties provide the simplest example of systematic 
uncertainties that can be represented by nuisance parameters in profile
likelihoods.  As an example, let us assume that the integrated luminosity
is measured in some auxiliary study, and results in a 2\% uncertainty.
We would rewrite the likelihood as
\begin{eqnarray}
    {\cal L} = \prod_{i=1}^N {\cal P}(n_i|\mu_i) {{\cal G}(L|\tilde{L},\sigma_L)}
\end{eqnarray}
for the measured value $\tilde{L}\pm\sigma_L$.  The function ${\cal
  G}$ is a normalized Gaussian of mean $\tilde{L}$ and width
$\sigma_L$, which serves to constrain the value of the new nuisance
parameter $L$ to its measured value.  Note that it is $L$ and not
$\tilde{L}$ that is used to calculate the $\mu_i$.  The negative log
likelihood is thus
\begin{eqnarray}
    -ln{\cal L} = \sum_i \left[-n_i\ln\mu_i + \mu_i\right] + \frac{(L-\tilde{L})^2}{2\sigma_L^2} 
\end{eqnarray}
and thus the remnant of the Gaussian term can be regarded as a penalty
on the negative log likelihood.  It is in principle possible to use
functions other than Gaussians to constrain the values of the nuisance
parameters.  In Bayesian terms the constraint functions are simply the
prior probability densities of the nuisance parameters.

Any multiplicative uncertainty can be represented in the likelihood 
this way, including uncertainties on cross sections, overall 
efficiencies, and the like.  One can also introduce multiplicative
nuisance parameters into Eq. 2 as needed, for any or all sources.

In many cases, however, the allowed physical bound on a multiplicative
nuisance parameter is that it remain positive.  If we are representing
the constraint by a Gaussian, then when the uncertainty in the
nuisance parameter is large the Gaussian is truncated and an
appropriate normalization factor should be included.  In such cases
one might also consider constraining the parameter with a log normal
or other probability density which does not allow the parameter to
vbecome negative.

\section{Shape Uncertainties and Morphing}

Many systematic uncertainties result in an overall distortion in the
shape of the observed spectrum.  A good example is an energy scale
uncertainty which affects all jet energies in an event in the same
direction.  If there are energy thresholds in the event selection,
changes in not only the shape but the overall normalization of the
efficiency (represented here by the $\epsilon_{ji}$ for source j in
bin i) as a function of the observables can result.

Such spectral distortions can be modeled by altering parameters (like
the energy scale) in the MC simulation and recalculating the
``shifted'' set of efficiencies.  If we were, for example, to raise
and lower the energy scale by one standard deviation, recalculating
the efficiencies, we would then have three measures of the shape (and
normalization) of the bin efficiency distribution, which we can call 
$\epsilon^-_{ji}$, $\epsilon^0_{ji}$, and $\epsilon^+_{ji}$.  Clearly
one can obtain more measures from other alterations of the energy scale,
though this can often be computationally very expensive.

We then face the question of how to turn our three measures of the
spectral shape into a continuous estimate in each bin as a function of
the energy scale factor.  To do this we introduce a ``morphing''
parameter which we will call $f$, and which is nominally zero (in the
case of no scale shift), and which has some uncertainty (usually Gaussian)
$\sigma_f$=1.

In this general technique, usually called ``vertical morphing'', we
might write the efficiency in a bin as a function of the morphing
parameter as
\begin{eqnarray}
    \epsilon_{ji} = \epsilon^0_{ji} + f \frac{\epsilon^+_{ji}-\epsilon^-_{ji}}{2} \ \ \ .
\end{eqnarray}
In this expression we see that we are treating the difference in the shifted
efficiencies in the bin as if they represent a measurement of the first-order
Taylor expansion around the nominal value.  This may or may not be a
reliable indicator of how the efficiency spectrum changes with energy scale.
Also note that for $f=\pm 1$ the above expression does not actually
yield $\epsilon^\pm_{ji}$!  

To provide a better estimate of the true behavior of the spectral
distortions we have introduced a technique whereby we interpolate
quadratically for $|f|<1$ and extrapolate linearly beyond that range.
This does result in the exact measured behavior of the spectrum at
$f=\pm 1$ but avoids large deviations from linear behavior outside the
range.  The value of the efficiency at any $|f|<1$ can be determined by
Lagrange interpolation:
\begin{eqnarray}
    \epsilon_{ji} = \frac{f(f-1)}{2}\epsilon^-_{ji}
                   -(f-1)(f+1)\epsilon^0_{ji}
                   +\frac{f(f+1)}{2}\epsilon^+_{ji}
\end{eqnarray}
Calculation of the linear extrapolation beyond this range is a
straightforward exercise for the reader.

Clearly if a more accurate representation of the morphing behavior is
required, one can, at the expense of computation and bookkeeping time
obtain additional shifted efficiency spectra and interpolate using a
higher order polynomial.  A good measure of whether this is a
worthwhile exercise is to examine the behavior of one's morphing
parameters as a function of the parameter of interest; if they tend to
go far from the sampled region (corresponding to one standard
deviation in the uncertainty) then it may be desirable to obtain more
measurements there, and parametrize the measured region with a higher
order polynomial.

We also note that there are somewhat more sophisticated methods such
as Alex Read's ``horizontal morphing''~\cite{Read} method.  These are
more computationally intensive, but could be advantageous.  However
they are not straightforwardly defined in more than one dimension.

The morphing method presented here can be extended to several morphing
parameters for different systematic effects simply by adding linearly
the deviations from the nominal efficiency due to each effect.

\section{Statistical Uncertainties in Efficiencies}

Typically one estimates the efficiency of each source in each bin
using a Monte Carlo simulation, and hence the statistical accuracy of
the estimate of the efficiency in each bin depends on the number of
MC events falling there.  Likewise, in other, possibly data-driven methods
for estimating the expected number of events from some source in some
bin, there may be some known statistical uncertainty in each bin.  

Barlow and Beeston~\cite{BB} proposed a method for representing such
systematic uncertainties wherein one introduces a separate nuisance
parameter multiplying the expected number of events from each source
in each bin.  Nominally the value of these parameters is 1, and one
can then constrain the parameters, which we call $\beta_{ji}$,
according to the prior pdf assumed for the number of MC events in the
efficiency calculation.  Barlow and Beeston assumed a Poisson
distribution (though one might argue a binomial is the most correct
form to assume); other choices such as log normal avoid the parameters
possibly tending to zero.

Though this method introduces a very large number of new free
parameters in the likelihood, the problem can be seen to be tractable
in the profile likelihood case since the values of the $\beta_{ji}$
which maximize the likelihood within a bin can be found independently
of those in all the other bins.  

Assuming a Gaussian constraint on the $\beta_{ji}$, we can
write the contribution to the negative log likelihood in a particular bin as
\begin{eqnarray}
   -ln {\cal L}_i = -n_i \ln(\sum_j\beta_{ji}\mu_{ji}) + \sum_j\beta_{ji}\mu_{ji} 
                    + \sum_j\frac{(\beta_{ji}-1)^2}{2\sigma_{ji}^2} \ \ \ .
\end{eqnarray}
This contribution can be extremized with respect to the $\beta_{ji}$
by setting the derivative with respect to each to zero.  Dropping the bin
index $i$ for clarity we write
\begin{eqnarray}
    \frac{\partial(-\ln{\cal L})}{\partial\beta_j}
      = \mu_j\left[1-\frac{n}{\sum_k\beta_k\mu_k}\right] + \frac{\beta_j-1}{\sigma_j^2} = 0 \ \ \ .
\end{eqnarray}
We thus arrive at a set of nonlinear equations for the $\beta_j$ in a bin.
These can be approximately solved by iterative Newton-type methods, or
by more sophisticated methods.

In the context of performing the profile likelihood using {\sc MINUIT}
minimization, one can implement this Barlow-Beeston type method by
solving for the $\beta_{ji}$ within the ``objective'' function which
provides to {\sc MINUIT} the value of $-\ln L$ given the values of all the
parameters in the fit, and include the contribution of the deviations
of the $\beta_{ji}$ from unity to $-\ln L$.

However, a problem arises in this approach.  Any minimization
algorithm can only approximate the values of the parameters and,
hence, the true minimum of a function.  There is always some last step
which meets the convergence criterion, and somewhere in the space of
the input $\mu_{ji}$ to the minimization for the $\beta_{ji}$, one
will find the place where that last step is not taken.  Near such
points the values of the resulting $\beta_{ji}$ and their associated
contribution to $-\ln L$ undergo a small discontinuous jump.  Such
jumps can (and do) dramatically confuse {\sc MINUIT}'s {\sc MIGRAD}
minimizer, which attempts to measure the Hessian matrix by finite
differences.  These jumps cause the resulting parameter covariance
matrix to become non-positive-definite.  When {\sc MINUIT} detects
such a situation it attempts to circumvent it by adding to the
offending diagonal element of the matrix an amount necessary to
restore positive-definiteness.  Sometimes this works but in many cases
all is lost: {\sc MINUIT} is now dealing with a false measure of the
Hessian matrix and it tends to send the free parameters in the fit to
wild values.  We have found no solution to this behavior short of
rewriting MINUIT.

The full-blown Barlow-Beeston method for dealing with bin statistical
uncertainties is not absolutely required to represent them properly in
the likelihood.  What matters, in a bin, is the {\em overall}
statistical uncertainty of the predicted number of events from all
sources.  The statistical uncertainties for each source in each bin
are independent, and can be readily combined, particularly if they are
Gaussian or Poisson in nature.  Thus, a single Barlow-Beeston type
parameter is sufficient to represent the statistical uncertainty.
The value of this parameter, and its contribution to $-\ln L$, can be
calculated exactly by solving a quadratic equation.  Using a
simplified notation for a single bin, we write
\begin{eqnarray}
    -\ln {\cal L} = -n\ln\beta\mu + \beta\mu + \frac{(\beta-1)^2}{2\sigma_\beta^2}
\end{eqnarray}
where here $\mu$ is the total number of expected events in the bin, given 
the values of all the other parameters, and $\sigma_\beta$ is the relative
(statistical) uncertainty in the prediction.  Setting the derivative to 
zero we find the quadratic equation
\begin{eqnarray}
    \beta^2 + (\mu\sigma_\beta^2 - 1)\beta -n\sigma_\beta^2 = 0
\end{eqnarray}
which can be solved readily and the correct root taken.  The extension to 
other constraint functions is straightforward though may result in 
transcendental equations to solve.

\section{Practical Considerations}

Care must be taken in using the approach described in this paper to avoid a
number of potential pitfalls which we discuss here.

\vspace{0.1in}
\noindent{\bf Sparsely Populated Bins}
\vspace{0.05in} 

In multi-bin spectra (particularly multi-dimensional spectra) one can
encounter situations where the number of events per bin varies by
orders of magnitude.  This can sometimes lead to situations where 
\begin{itemize}
   \item there can be regions of zero-content bins, surrounded by bins
         populated by single MC events;
   \item such single MC-event-bins can migrate under the influence of the 
         morphing systematic effects, spoiling the vertical morphing method;
   \item single data events can appear in bins where there is no predicted
         rate.
\end{itemize}
All of these situations must be avoided.  The most straightforward is to
generate sufficient Monte Carlo in all bins, but this may not be
practical or even possible.  The best alternative is to combine bins 
according to some algorithm (which does not use the observed data 
distribution!) which ensures some minimum statistical threshold in 
every bin in the fit.

\vspace{0.1in}
\noindent{\bf Bins Entering/Leaving the Likelihood}
\vspace{0.05in} 

It is also necessary to ensure that no bin enters or leaves the
likelihood as the parameters change.  It is not impossible for {\sc
  MINUIT} to drive parameters to regions where the contribution from a
source, or even all sources, vanishes in a bin.  For example, when
studying the profile likelihood as a function of some new particle
signal, also, one in general wants to evaluate the likelihood for the
case of zero signal.  But if there are bins populated by signal only,
this can cause the contribution to go to zero, the logarithm of which
is of course $-\infty$.

Simply excluding bins from the likelihood when there are no expected events
is not a sufficient solution to this problem, as a moment's reflection
will make clear.  To avoid bins entering/leaving the fit, therefore, 
the bins to be used or not used must be established {\em a priori} by
finding all bins where some contribution is expected, and making sure there
are no bins with data but no expected contribution.  Once determined, this
set must remain fixed for the duration of the calculation.

One way to ensure that no bin leaves the calculation is to always have
it contribute at least some tiny amount.  For example to circumvent
the zero-signal issue, we always ensure that the signal cross section
is no less than $10^{-10}$ pb, and that no source in any bin used in
the fit ever contributes less than $10^{-10}$ expected events.  Though
this is a somewhat inelegant solution to a nevertheless important
problem, we note that our final results do not depend on these minimum
values in practice.

\section{Pseudo-Bayesian Posterior Densities}

For measuring physical parameters, the profile likelihood can be
directly interpreted using the usual $\Delta(\ln L)$ approach to
derive confidence intervals in multi-dimensional parameter space.

To extend this treatment to setting exclusion bounds on parameters
such as a hypothetical new particle's cross section $\sigma_X$, we can
simply derive a posterior density by treating the profile likelihood,
which we shall denote ${\cal L}_{max}$ as one would any likelihood
using Bayes' Theorem:
\begin{eqnarray}
       {\cal P}(\sigma_X) = \frac{{\cal L}_{max}(\sigma_X){\cal P}(\sigma_X)}
                                  {\int_0^\infty {\cal L}_{max}(\sigma_X){\cal P}(\sigma_X) d\sigma_X}
\end{eqnarray}
where here ${\cal P}(\sigma_X)$ is the assumed prior pdf in
$\sigma_X$.\footnote{All the usual inveighments against improper
priors apply at this point.  We would like to point out, however, that 
in every case of which we are aware, where such a posterior is used to 
quote confidence intervals on the parameter of interest in an actual
{\em measurement} of that parameter, no one ever uses a prior other than 
a uniform one.}

But does the profile method really result in a posterior density that
can be interpreted in this way?  The most proper Bayesian treatment
would not maximize the likelihood with respect to the parameters not
of interest, but marginalize instead, resulting in what we might
denote as $\bar{\cal L}(\sigma_X)$ to highlight the fact that the
marginalized likelihood is in a sense averaged over the prior-weighted
values of the nuisance parameters.  

We have performed both calculations, profiling and marginalization, in
a variety of complex spectrum fits, and it is our experience that the
posterior density derived either way is nearly identical, though the 
marginalized one takes orders of magnitude more compute time.  Due to this 
practical consideration alone we employ the profile method and consider 
it to be a near-perfect representation of a full and proper Bayesian 
marginalization treatment.

\section{Conclusions}

We present in this paper the basic mathematical and numerical approach
to fitting multi-source spectra using a profile likelihood in which
various types of systematic uncertainties are incorporated by
representing them by nuisance parameters.  This method, we believe,
offers a unified approach to setting exclusion bounds, making
discoveries, and ultimately performing measurements on a wide range of
particle physics data analyses.

\section*{Acknowledgements}

I wish to thank my CDF $t^\prime$ colleagues Andrew Ivanov, Robin
Erbacher, Alison Lister, David Cox, Will Johnson, and Thomas Schwarz
for their insights and ideas in developing these methods.  This work was
supported by the US Department of Energy Office of Science.

\end{document}